\documentclass[seceq]{ptptex}
\usepackage{graphicx}
\usepackage{color}

\definecolor{Red}{rgb}{1.00, 0.00, 0.00}
\newcommand{\red}{\color{Red}}
\definecolor{Green}{rgb}{0.0, 1.0, 0.0}
\newcommand{\green}{\color{Green}}

\newcommand{\bitem}{\begin{itemize}}
\newcommand{\eitem}{\end{itemize}}
 \newcommand{\bmini}{\begin{minipage}}
 \newcommand{\emini}{\end{minipage}}
\newcommand{\be}{\begin{eqnarray}}
\newcommand{\ee}{\end{eqnarray}}

\newcommand\nn{\nonumber}

\newcommand{\mat}{\left ( \begin{array}{cc}}
\newcommand{\emat}{\end{array} \right )}
 \newcommand{\vect}{\left ( \begin{array}{c}}
\newcommand{\evect}{\end{array} \right )}

\newcommand{\colgy}{}
 \newcommand{\bflus}{\begin{flushright}}
\newcommand{\eflus}{\end{flushright}}

\nofigureboxrule                     
\notypesetlogo                       
\title{
Random Matrix Theory at Nonzero $\mu$ and $T$
}

\author{
Kim  \textsc{Splittorff}$^{1,}$\footnote{ e-mail address:
split@nbi.dk}  
and
Jacobus Johannes Maria \textsc{Verbaarschot}$^{1,\,2\,,3\,,} $\footnote{ e-mail address:
jacobus.verbaarschot@stonybrook.edu} 
}

\inst{
$^1$ 
The Niels Bohr Institute, Blegdamsvej 17, DK-2100, Copenhagen {\O}, Denmark\\
$^{2}$
Niels Bohr International Academy, Blegdamsvej 17, DK-2100, Copenhagen {\O}\\
$^3$
Department of Physics and Astronomy, SUNY, Stony Brook,
 New York 11794
}

\abst{
We review applications of random matrix theory to QCD at nonzero 
temperature and chemical potential. The chiral phase transition of
QCD and QCD-like theories is discussed in terms of eigenvalues of the
Dirac operator. We show that for QCD at $\mu \ne 0$, which has a sign problem,
the discontinuity in the chiral condensate is due to an alternative to the 
Banks-Casher relation. The severity of the sign problem is analyzed in the 
microscopic domain of QCD.
}

\begin{document}

\maketitle

\section{Introduction}

Starting from its introduction in nuclear physics by Wigner \cite{wigner},
random matrix theories have been applied to a wide range of
problems ranging from the physics of proteins \cite{melih}
to quantum gravity (see \cite{m2,snaith} for a historical review). 
Three reasons for the ubiquity of 
random matrix theory come to  mind.
First, eigenvalues of large random matrices
have universal properties determined by symmetries.
Second, random matrices are models for disorder present in many
physical systems. Third, random matrix theories have a topological expansion
which is important for applications to quantum field theory.
One of the attractive features of random matrix theory 
is that analytical information
can be obtained  for complex systems which otherwise only can be studied
experimentally or numerically. 

In this review we discuss applications of random matrix theory to QCD
at nonzero temperature and chemical potential. Since the 
order parameter
for the chiral phase transition \cite{BC,OSV} and the deconfining 
phase transition \cite{Gattringer,wipf} are determined by the
infrared behavior of the eigenvalues of the Dirac operator, these 
eigenvalues are essential for the  phase
transitions in QCD. Remarkably, the distribution of the smallest
Dirac eigenvalues is given by universal functions 
\cite{SV,V,VZ,NDW,DN} that depend
only on one or two parameters, the chiral condensate and the pion 
decay constant. This offers an alternative way to measure these
constants on the lattice
\cite{Berbenni,DHNR,DeGrand,Fukaya,Lang,heller,heller2,OW,kim}.

\section{ Random Matrix Theory in QCD}

Chiral Random Matrix Theory (chRMT)
is a theory with the global symmetries of QCD, but matrix elements
of the Dirac operator replaced by random numbers \cite{SV,V}
\be
D = \mat m & iW \\ i W^\dagger & m \emat, \quad P(W) \sim 
e^{-N {\rm Tr} W^\dagger  W}.
\label{chrmt}
\ee 
This random matrix model  has the global symmetries 
and topological properties of QCD. It is confining in the sense
that only color singlets have a nonzero expectation value.
It is now well understood that fluctuations of 
low-lying eigenvalues of the Dirac operator are described 
by chRMT (see \cite{ency,tilorev,nowak,houches,splitrev,gerrev} for lectures 
and reviews).
Philosphically, this is important because of the realization 
that chaotic motion dominates the dynamics of quarks at low energy.
Practically, this is important because we can
use powerful random matrix techniques to calculate physical
observables.

The condition for the applicability of chRMT
is that the Compton wavelength of Goldstone
bosons associated with the mass scale $z$ of these eigenvalues is much
larger than the size of the box. With the squared mass of the associated
Goldstone boson given by $2 z \Sigma/F_\pi^2$, this condition reads \cite{Vplb}
\be
\frac {2z \Sigma}{F^2_\pi} \ll  \frac 1{\sqrt V} \ll \Lambda^2 .
\label{domain}
\ee
The second condition is necessary to factorize the partition function
into a contribution from the lightest degrees of freedom and all
heavier degrees of freedom. These two conditions determine the microscopic
domain of QCD.
We stress that $z$ is a scale in the Dirac
spectrum so that, for sufficiently large volumes,  we  always have 
eigenvalues in the domain (\ref{domain}) where eigenvalues
fluctuate according to chRMT. This can be
shown rigorously from the following two observations \cite{OTV,DOTV}.
First, the infrared Dirac spectrum follows from
a (partially quenched) chiral Lagrangian determined by chiral symmetry, and
the inequality (\ref{domain}) is the condition for factorization of the
partition function into a factor containing the constant modes
and another factor containing the nonzero momentum modes. 
Second, the factor with the constant modes is equal to the large
$N$ limit of chiral random matrix theory.

In \cite{GLeps,LS} the condition (\ref{domain}) 
 was imposed on the quark masses and was the bases for 
a systematic expansion of the chiral Lagrangian 
known as the $\epsilon$ expansion. 

One feature that underlies universal properties of eigenvalues is
that they behave as repulsive
confined charges. This follows from the joint probability distribution
$\sim  \prod_k \lambda_k 
\prod_{k<l} (\lambda_k^2-\lambda_l^2)^2 \exp(-N \sum_k \lambda_k^2)$.
It can be shown that eigenvalues correlations at the micrsocopic  
scale are universal \cite{ADMN}. 
The reason is spontaneous symmetry breaking and a mass gap so that 
they can be described in terms of a  chiral Lagrangian.

\subsection{Chiral Random Matrix Theory at $\mu\ne 0$ and $T\ne 0$}

A nonzero temperature does not change the fluctuating
behavior of the Dirac eigenvalues provided that chiral symmetry
remains broken. However, a transition to a different universality class
takes place at the critical temperature. A random
matrix model that reproduces this universal behavior of QCD
is obtained by replacing the off-diagonal elements
in (\ref{chrmt}) by \cite{JV}
\be
iW \to iW + t,\qquad iW^\dagger \to iW^\dagger - t \qquad {\rm with} 
\quad t = {\rm diag} ( -\pi T,\pi T ).
\label{ojv}
\ee
This model has been studied elaborately in the literature (see e.g. 
\cite{JV,tilo-t,misha-t,melih-u,papp,janik}).

A nonzero chemical potential can be introduced analogously to the
quark mass. The requirement is that the small $\mu$ behaviour of 
the QCD partition function should be reproduced by the random matrix 
partition function. This achieved by modifying 
(\ref{chrmt}) by \cite{misha}
\be
iW  \to iW +\mu, \qquad iW^\dagger \to iW^\dagger +\mu,
\label{mrmt}
\ee
resulting in a nonhermitean Dirac operator with eigenvalues scattered 
in the complex plane. The prescription (\ref{mrmt}) is not unique. 
A random matrix model that has had a strong impact on recent
developments is defined by \cite{O}
\be
iW  \to  iW + \mu H, \qquad iW^\dagger  \to  iW^\dagger + \mu H \qquad 
{\rm with} \quad H^\dagger = H ,
\label{ormt}
\ee
where $H$ is drawn from a Gaussian ensemble of random matrices. This
model is in the  
same universality class as (\ref{mrmt}) but is technically simpler
since it can be worked out by means of the complex 
orthogonal polynomial method \cite{fyodorov,ak-po,BI,O,AP}.
\begin{figure}[t!]
\begin{center}
\includegraphics[width=4.5cm]{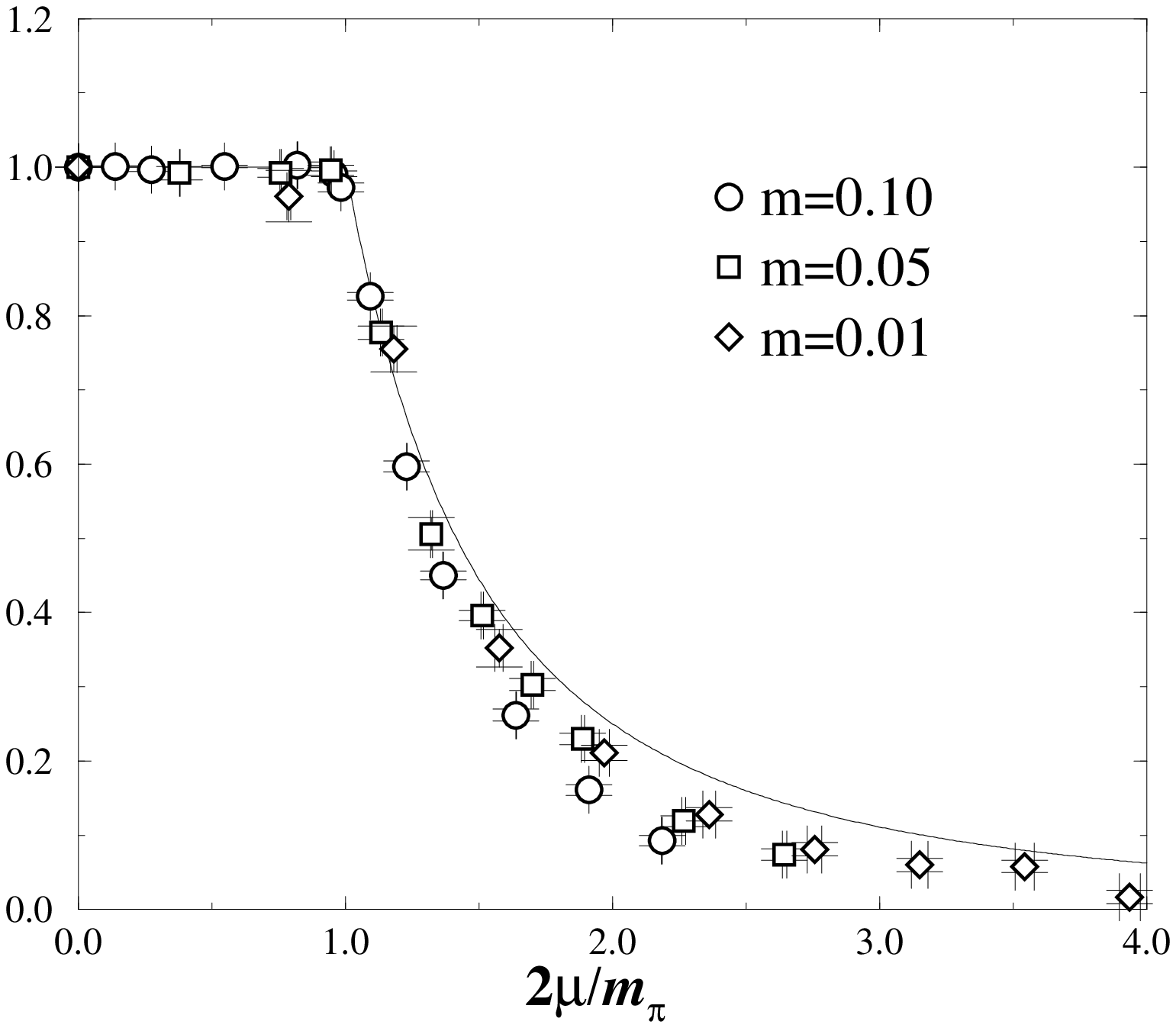}
\includegraphics[width=4.5cm]{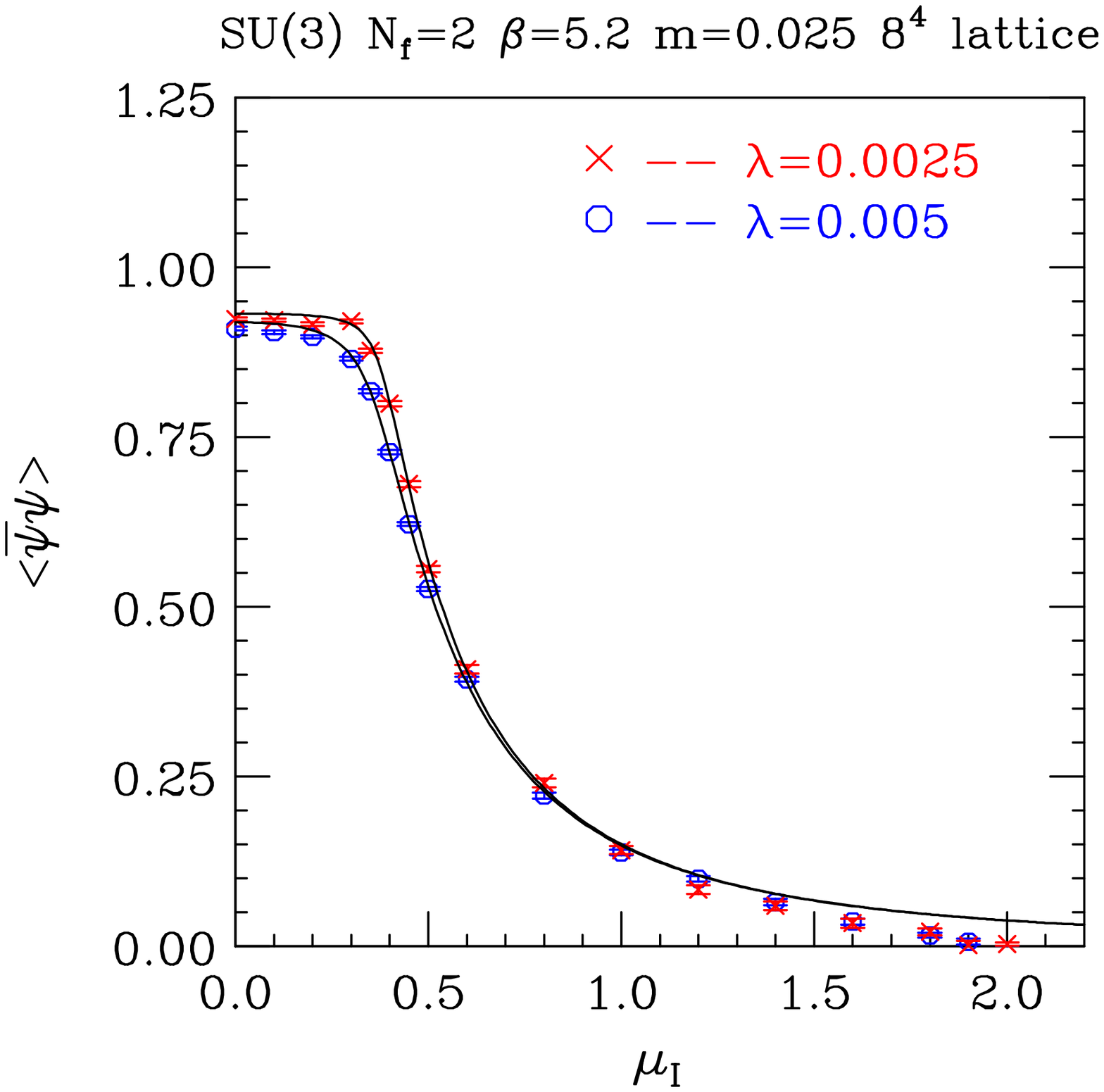}
\end{center} 
\caption{Lattice results for $N_c=2$ (taken from \cite{hands}) and
phase quenched QCD with $N_c=3$ (taken from \cite{sinclair})}
\label{fig1}
\end{figure}

There are other types of random matrix models that have been applied
to QCD. For example models with random gauge fields
such as the Eguchi-Kawai model
 \cite{Eguchi} or its 2-dimensional version \cite{GW}. QCD in 1 dimension
\cite{gibbs1,bilic} is a random matrix model as well, with universally
fluctuating Dirac eigenvalues. Also models with random Wilson loops
\cite{olesen,pisarski} have attracted significant interest.
\section{Phases of QCD and RMT}

QCD-like theories with charged Goldstone bosons have a critical chemical
potential equal to $m_\pi/2$. The phase transition to
the Bose condensed phase can therefore be described completely in terms of 
a chiral Lagragian. At the mean field level \cite{KSTVZ}, the kinetic
terms of this chiral Lagrangian do not contribute, so that these 
results can also be obtained from chiral  random matrix theory.
 Indeed, the static part
of the chiral Lagrangian \cite{KST,KSTVZ}
\be 
{\cal L} = \frac 14 F_\pi^2 \mu^2 {\rm Tr} [U,B][U^\dagger,B] -
\frac 12 \Sigma {\rm Tr} (M U + M U^\dagger).
\label{lmu}
\ee
can also be obtained from the large N limit of the models (\ref{mrmt})
or (\ref{ormt}).

In Fig. \ref{fig1} we display lattice results for QCD with $N_c=2$ \cite{hands}
and phase quenched QCD \cite{sinclair}. They show an impressive agreement
with the results from (\ref{lmu}) given by the solid curves in both 
figures.

\subsection{Schematic RMT Phase Diagram}

\begin{figure}[t!]
\begin{center}
\includegraphics[width=2.5cm]{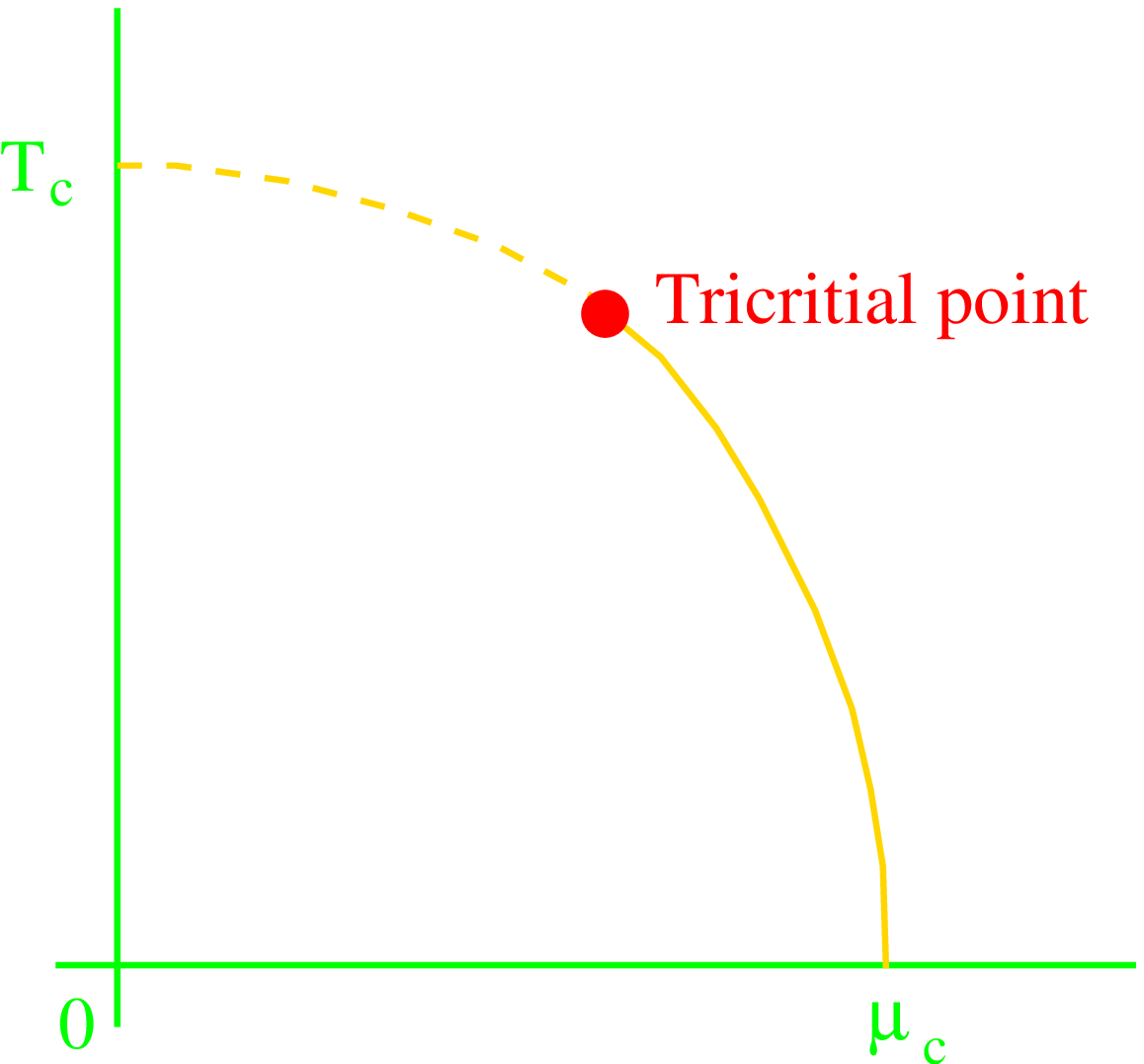}\hspace*{2cm}
\includegraphics[width=6cm]{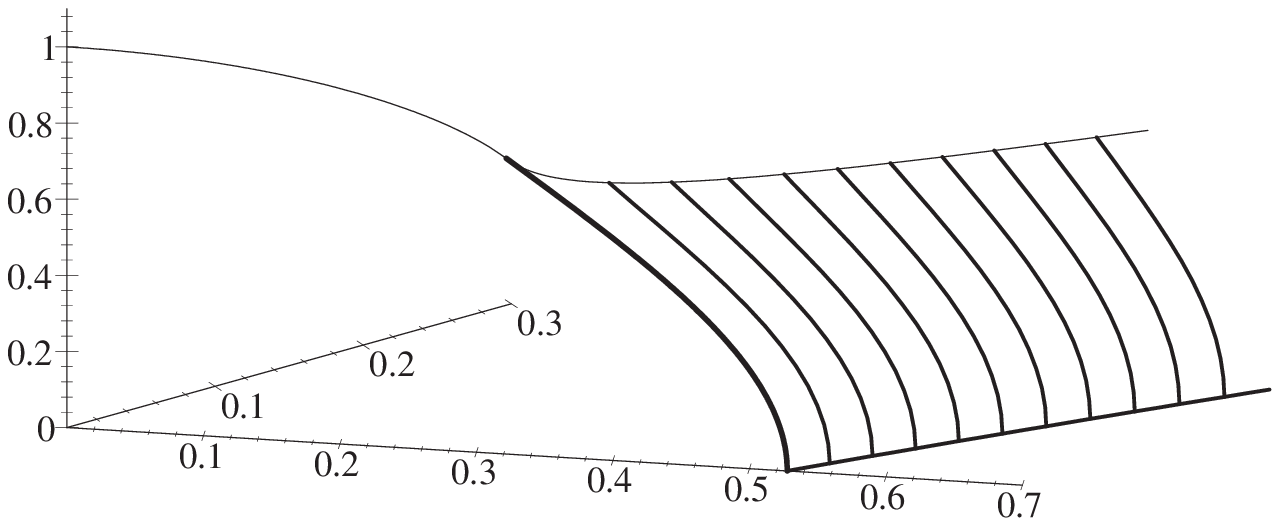}
\caption{QCD phase diagram in the $\mu T m$-space
(taken from \cite{tri})
}
\end{center}
\label{fig2}
\end{figure}

The phase transition in  QCD with $N_c=3$  
at $\mu_c =m_N/3$  cannot be analyzed by means of chiral 
Lagrangians. Because of the sign problem lattice studies are 
not possible either. In such situation there
is long tradition to analyze the same problem in a much simpler
theory in the hope of obtaining at least a qualitative understanding
of the problem. For example, one dimensional QCD  \cite{gibbs1,bilic}, 
or more recently, super Yang-Mills
theory and AdS-CFT duality \cite{son},  been explored as toy models for 
QCD.  
We will use random matrix theory at $T\ne0$ and $\mu\ne0$,
introduced in (\ref{ojv}) and (\ref{mrmt}) to obtain
a qualitive understanding of the QCD phase diagram. 
Lattice QCD simulations show that the chiral phase transition
at $\mu = 0$ is of second order or a steep cross-over. At $T=0$ we
expect a first order phase transition at $\mu_c=m_N/3$. 
It is natural that the
first order line ends in a critical end point or joins the second
order critical line at the tricritical point (see Fig. \ref{fig2}, left).
This is indeed what is observed in random matrix theory \cite{tri,benoit}
(see Fig. \ref{fig2}, right).
A similar phase diagram has also been obtained from 
the NJL model \cite{barducci,berges,njl0}.

Another scenario that was discovered in RMT is the splitting of 
the first order line into two
at nonzero isospin chemical potential \cite{klein}. This behavior was
also found in a NJL model \cite{lorenzo,walters} but might
not be stable against flavor mixing interactions \cite{buballa}.

\section{ Dirac Spectrum in Theories Without a Sign Problem}

Since the spectrum of the Dirac operator determines
the chiral condensate,  phase
transitions in QCD can be understood in terms of its spectral 
flow.  In this section we discuss
theories with a positive fermion determinant 
such as QCD with two colors and phase quenched QCD, where 
 a probabilistic
interpretation of the eigenvalue density is possible. The relation
between chiral symmetry breaking and Dirac spectra is much more
complicated when the fermion determinant is complex and its discussion
will be postponed to the next section.

The spectrum of an anti-Hermitean Dirac operator is purely
imaginary with an eigenvalue density that is proportional to the volume.
If chiral symmetry is broken spontaneously, 
the chiral condensate
becomes discontinuous across the imaginary 
axis in the thermodynamic limit. 
Chiral symmetry is restored if such discontinuity is absent
for example by the formation of a gap  in the Dirac 
spectrum, see eg.\cite{Far}\, . 

\begin{figure}[t!]
\includegraphics[width=14cm]{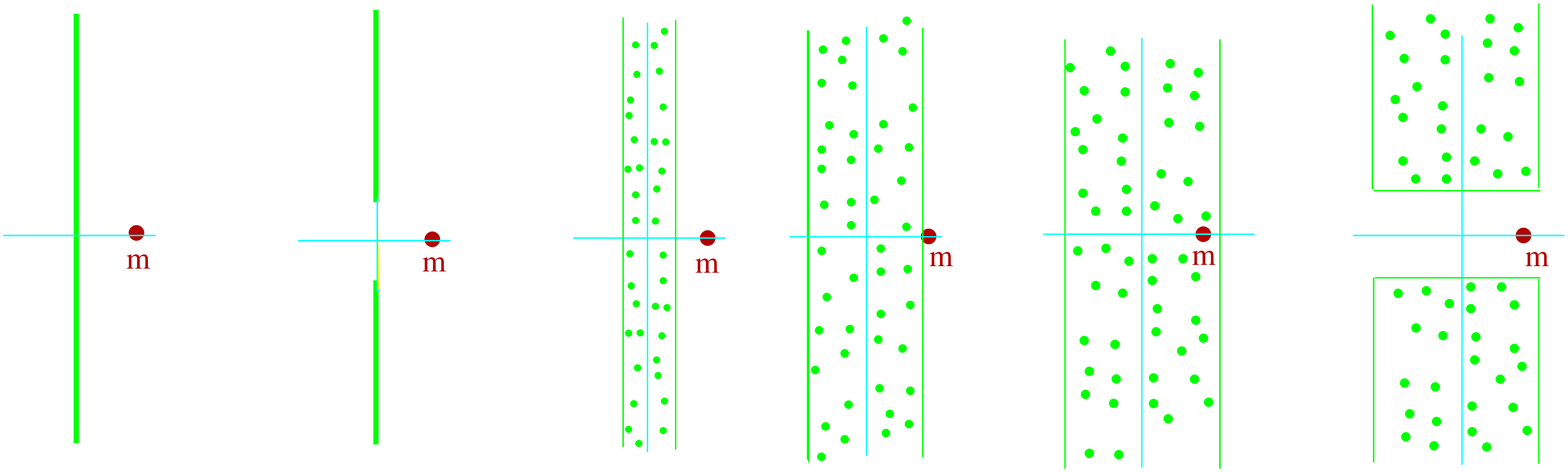}\\
\put(0.1,0.2){\tiny \bmini{2cm}$T<T_c$\\  $\mu= 0$ \emini}
\put(2.9,0.2){\tiny \bmini{2cm}$T>T_c$\\  $\mu= 0$ \emini}
\put(5.0,-0.0){\tiny \bmini{2cm}$T<T_c$\\  $\mu<\mu_c$ \emini}
\put(7.4,-0.0){\tiny \bmini{2cm}$T<T_c$\\  $\mu=\mu_c$ \emini}
\put(9.6,-0.0){\tiny \bmini{2cm}$T<T_c$\\  $\mu>\mu_c$ \emini}\
\put(12.4,-0.0){\tiny \bmini{2cm}$T>T_c$\\  $\mu>\mu_c$ \emini}
\caption{Critical behavior of the Dirac spectrum. $\mu_c=m_\pi/2$
  for $T=0$ and increases with $T$.}
\label{fig3}
\end{figure}

\begin{figure}[b!]
\begin{center}
\includegraphics[width=3cm]{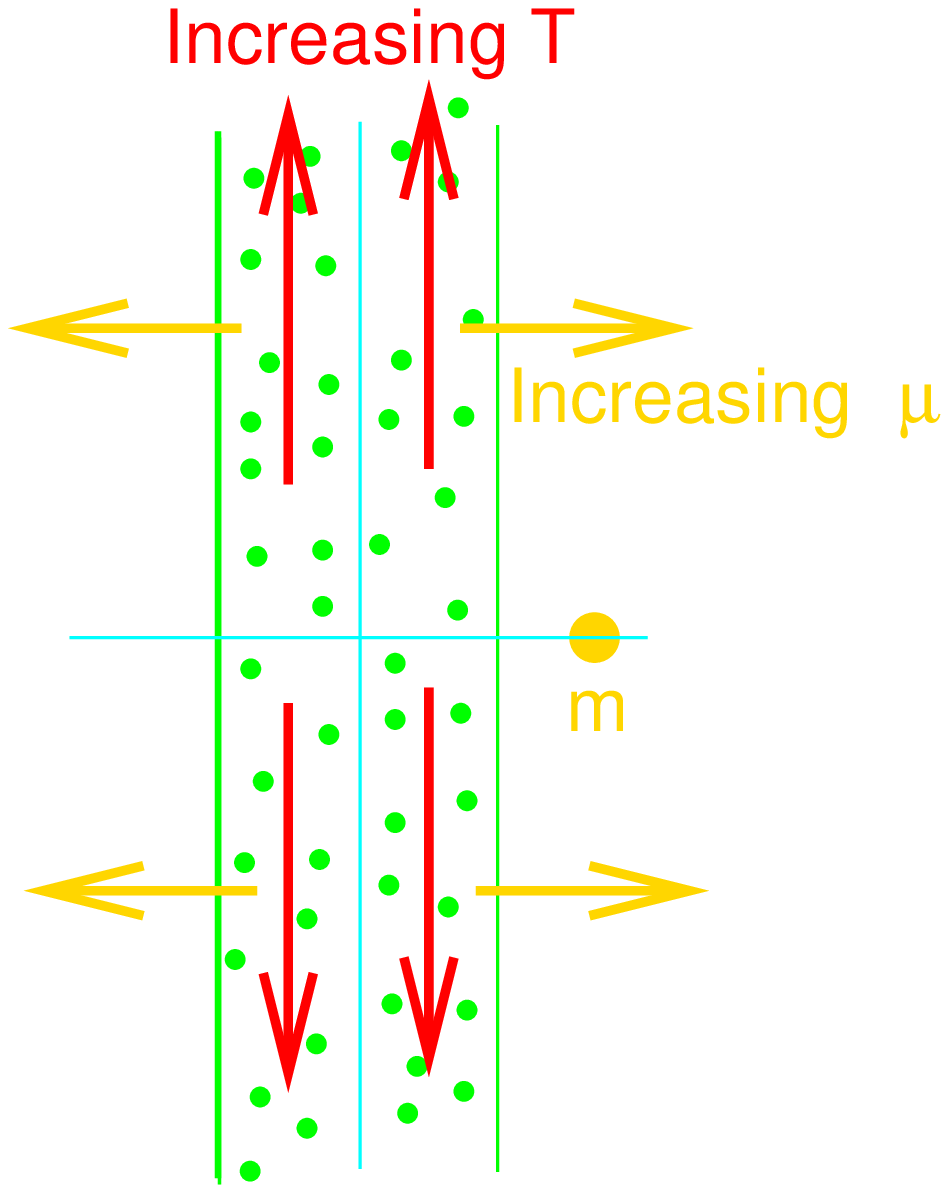}
\hspace*{2cm}\includegraphics[width=3cm]{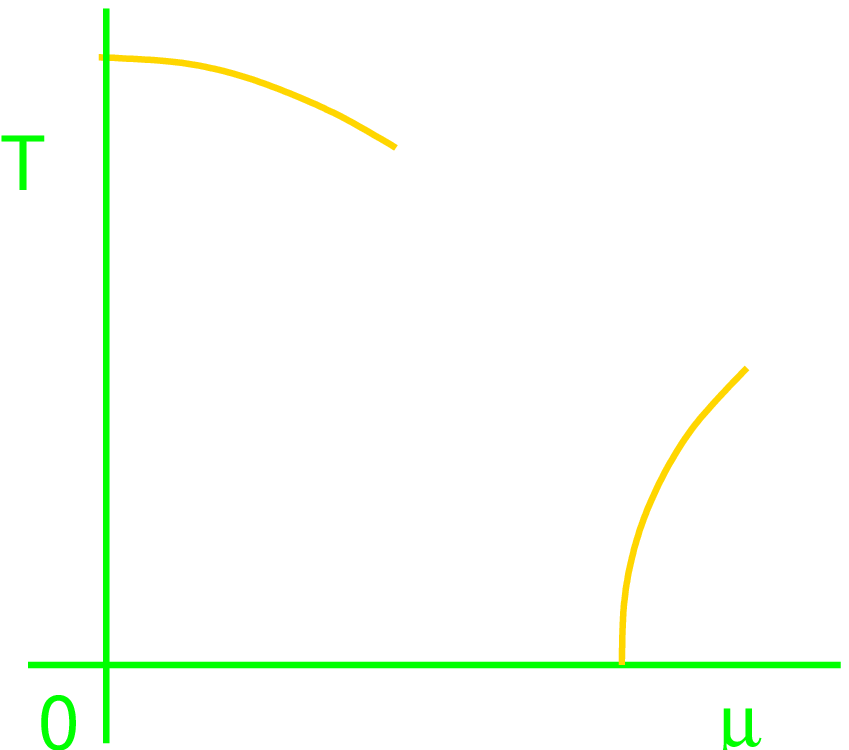}
\end{center}
\label{fig4}
\caption{Spectral flow of the Dirac spectrum (left) and phase diagram
  (right) with respect to $\mu$ and $T$ in phase quenched QCD and QCD
  with two colors.}
\end{figure}

\begin{figure}[t!]
 \includegraphics[width=4.cm]{lowests.eps}
\vspace*{-0.3cm}
{\includegraphics[width=4cm]{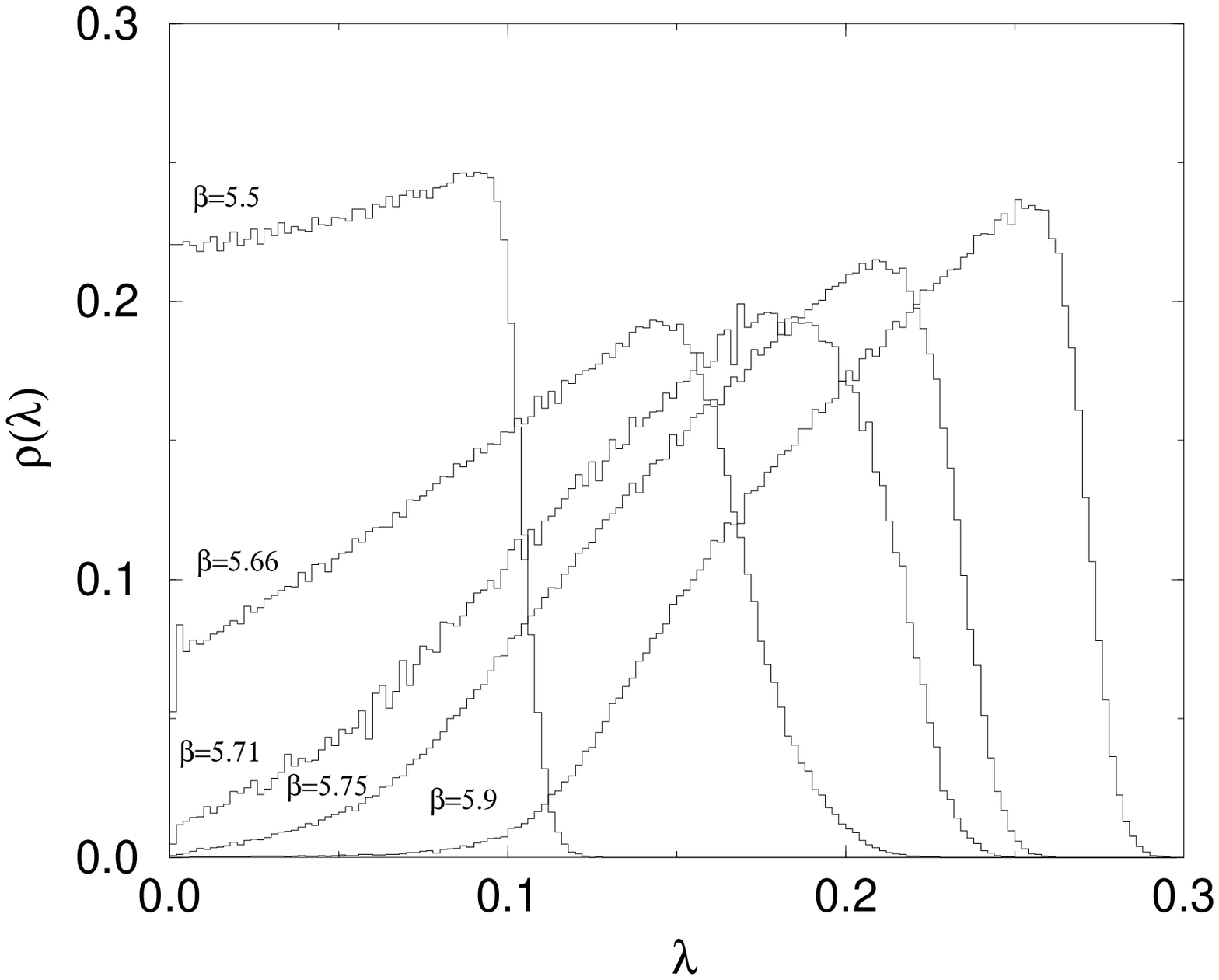}}
\vspace*{0.3cm}
 \includegraphics[width=2.5cm]{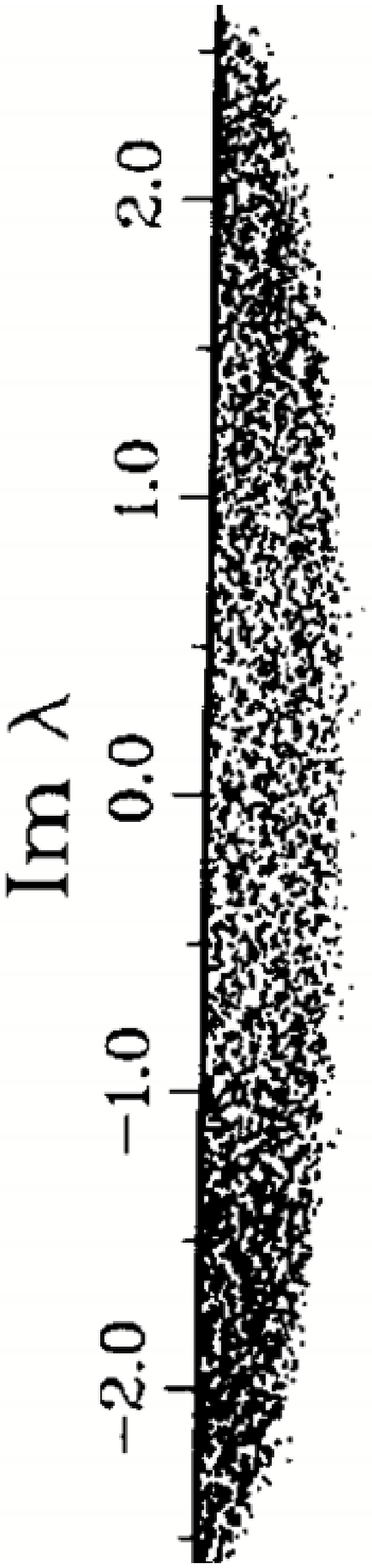}
\includegraphics[width=2.5cm]{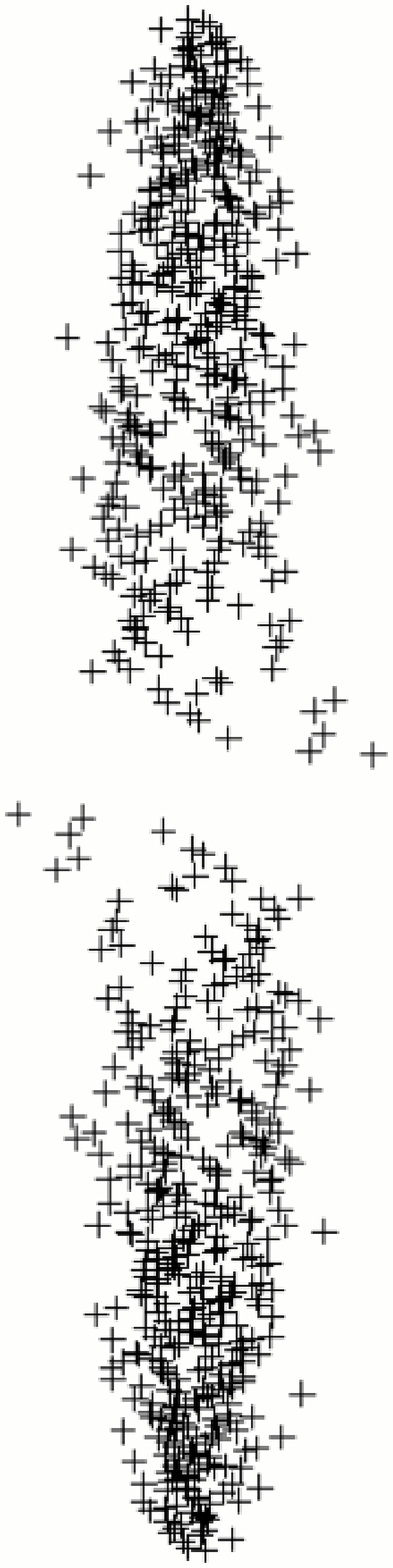}
\caption{Temperature and chemical potential dependence of Dirac 
eigenvalues. From left to right taken from 
\cite{narayanan,Damg,early,Muroya}.}
\label{fig5}
\end{figure}

For $\mu\ne0$, the Dirac spectrum broadens into
a strip of width $4\mu^2 F^2_\pi/\Sigma$ \cite{gibbs1,TV}. The chemical
potential becomes critical when the quark mass hits the
edge of this strip. At this point the chiral condensate starts
rotating into a pion condensate. Chiral symmetry restoration takes  place
when a gap forms at zero. A schematic picture of the critical behavior of
Dirac eigenvalues is shown in Fig. \ref{fig3} and
the spectral flow of the Dirac eigenvalues with respect to
increasing $\mu$ and $T$ is summarized in Fig. \ref{fig4}.
\setlength{\unitlength}{1cm}
One conclusion from this behavior is that
{$T_c(\mu)$ } is a concave function of {\colgy $\mu$}, and that
{ $\mu_c(T)$} is a convex function of {\colgy $T$}. 
The spectral flow discussed in this section is supported by lattice
simulations  at $T \ne 0$ and $ \mu \ne 0$ (See Fig. \ref{fig5})

\subsection{Dirac spectrum in the $\mu$-plane}

\begin{figure}[b!]

\includegraphics[width=3.5cm]{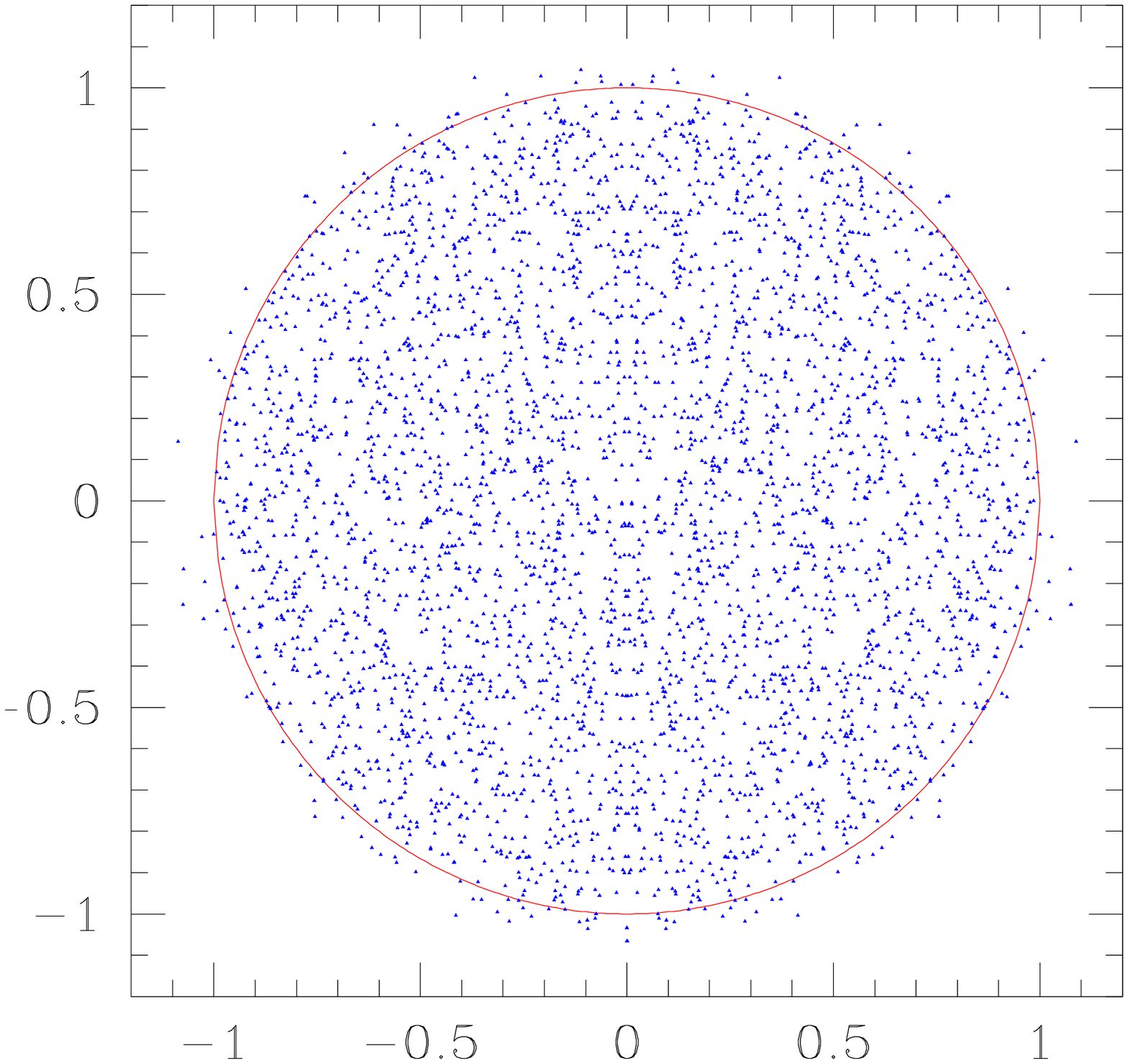}\hspace*{1cm}
\includegraphics[width=3.5cm]{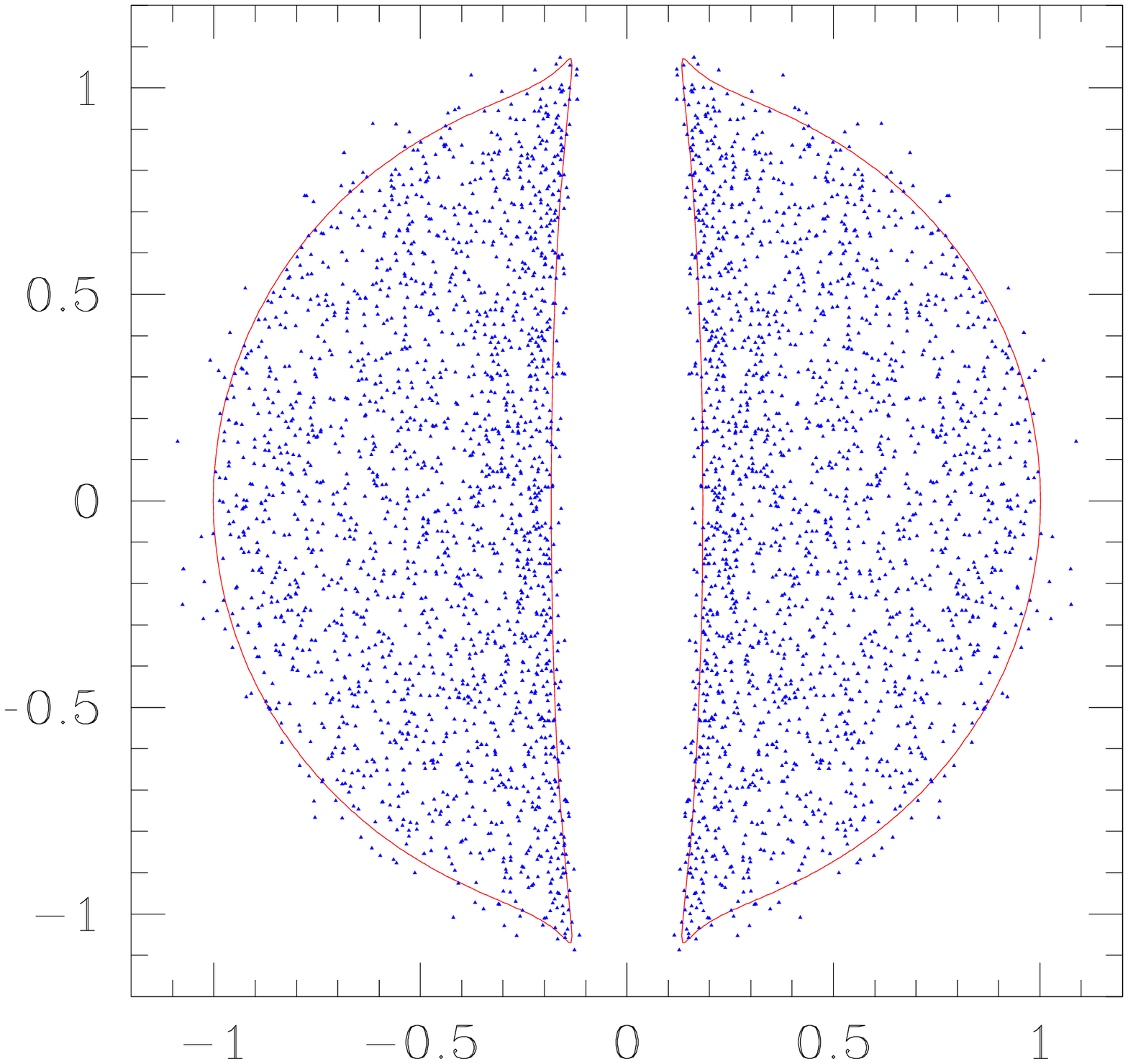}\hspace*{1cm}
\includegraphics[width=3.5cm,angle=90]{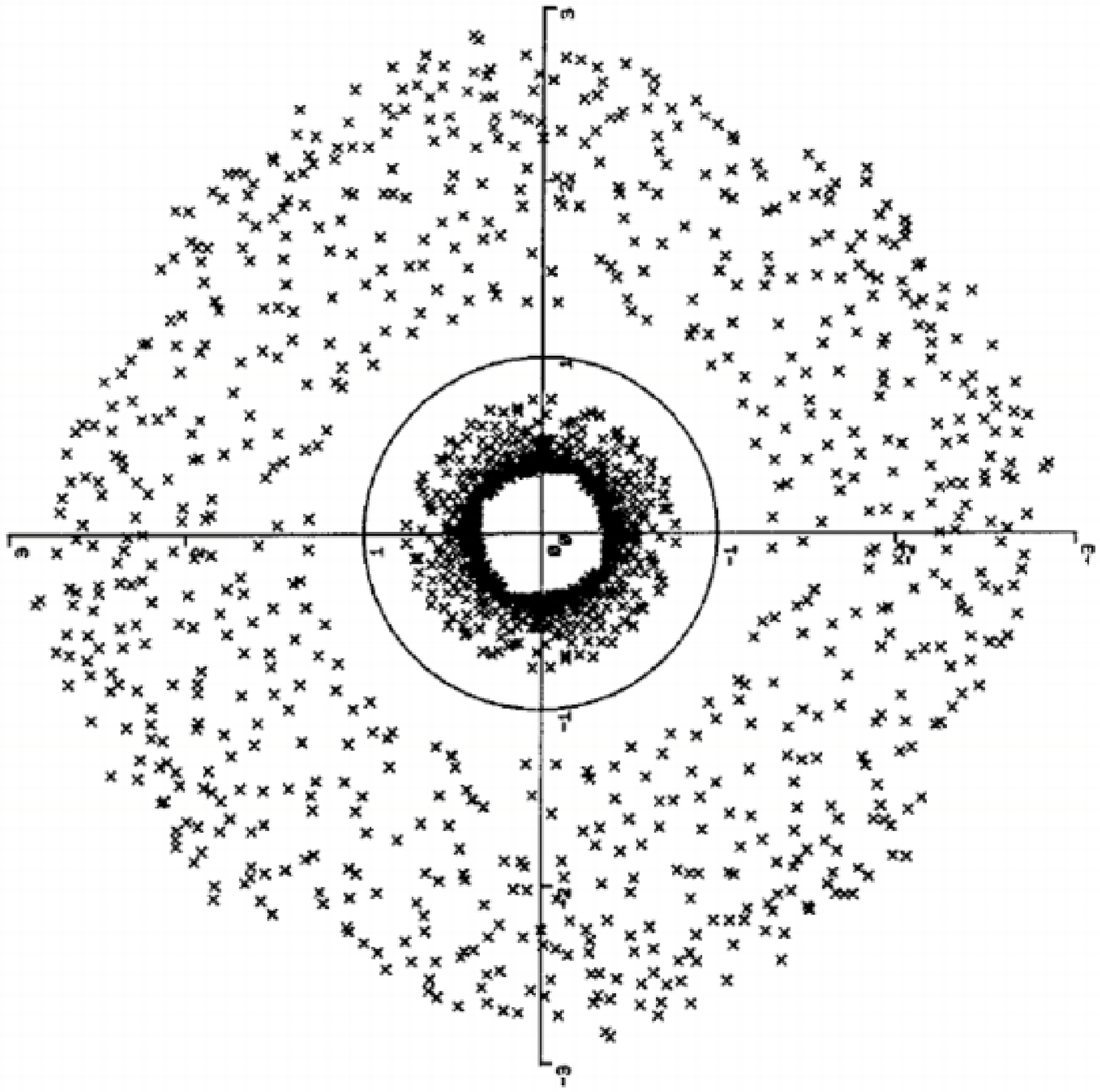}\\
\begin{picture}(10,.1)  
  \put(2.1,2.0){\red$\bullet \atop {\bf \mu}$}
  \put(6.7,2.0){\red $\bullet \atop {\bf \mu }$}
\end{picture}
\caption{Eigenvalues of   $\gamma_0(D+m)$ for a random
matrix Dirac operator at $m=0$ (left), $m\ne 0$ (middle) 
(both taken from \cite{HSV}),
and
lattice QCD at $m \ne 0$ (right, taken from \cite{gibbs1}).}
\label{fig6}
\end{figure}

We could equally well have diagonalized the Dirac operator in a representation
where {\colgy $\mu\gamma_0$} is proportional to the identity,
\be
\det(D+m + \mu \gamma_0) = \det(\gamma_0(D+m) + \mu).
\ee
These eigenvalues are relevant  to the baryon number density.
 A gap in the spectrum develops at $m\ne 0$ (see Fig. \ref{fig6}), and the
chemical potential becomes critical, $\mu=m_\pi/2$ when it hits the
inner edge of the domain of eigenvalues.

\subsection{Quenched Lattice QCD Dirac Spectra at $\mu\ne 0$}
 
\begin{figure}[t!]
\includegraphics[width=4.5cm,angle=0]{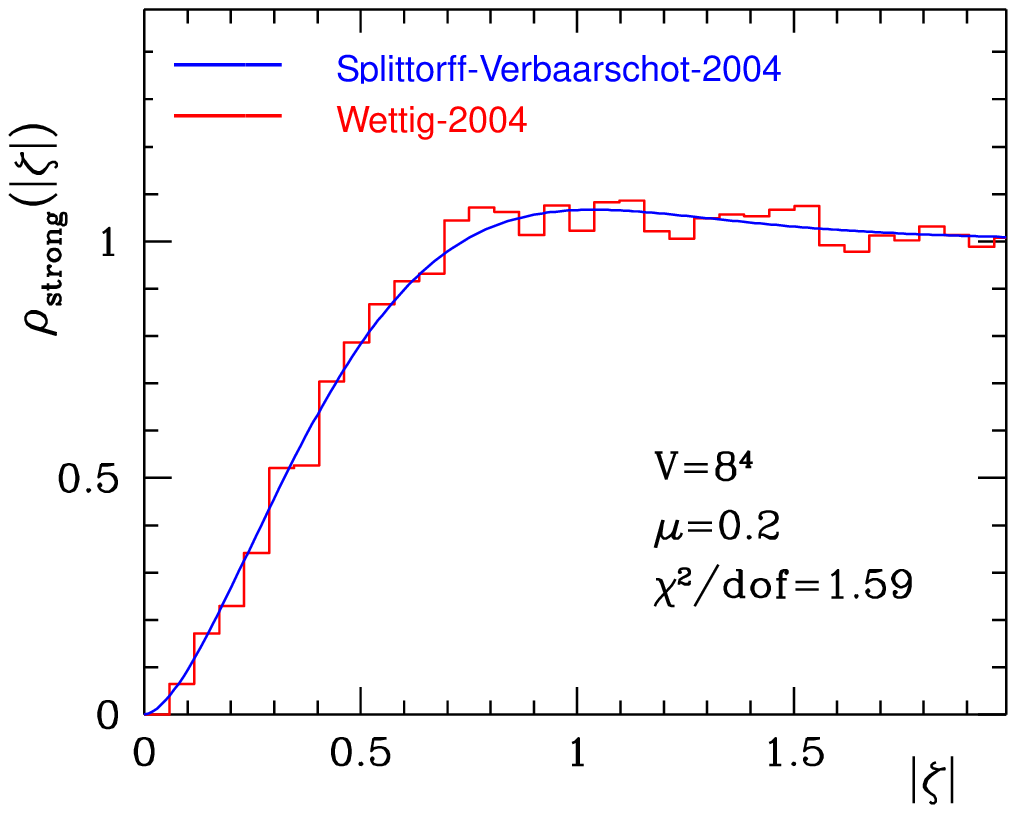}\hspace*{1cm}
\includegraphics[width=4cm]{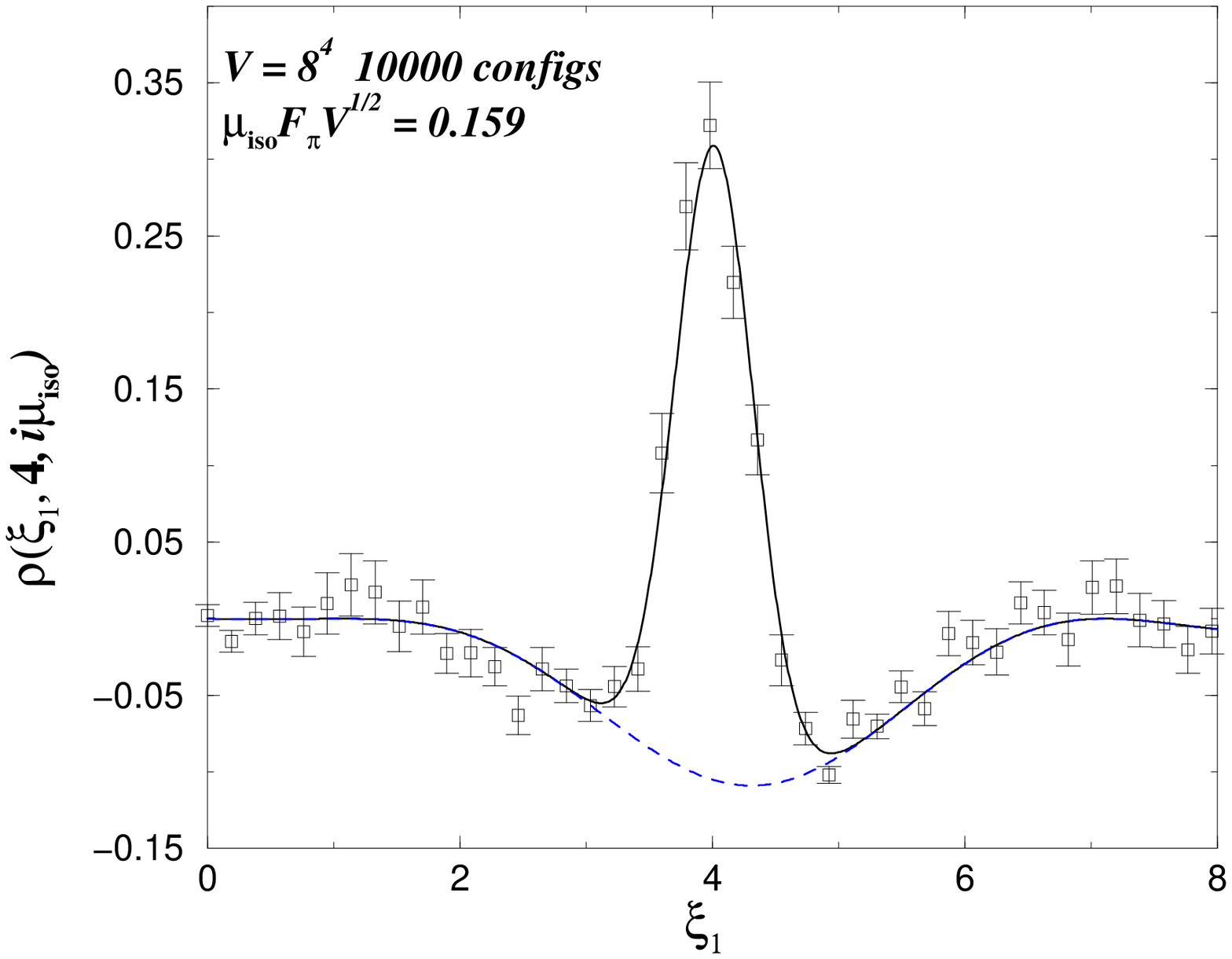}\hspace*{1cm}
\includegraphics[width=3.5cm]{dt5061.eps}
\caption{The radial spectral density for    
(left, taken from \cite{tilopr,tilo-ger}) and two-point correlations
(middle taken from \cite{heller} and right taken from \cite{OW}).}
\label{fig7}
\end{figure}

Small Dirac eigenvalues at $\mu \ne 0$ have been computed in quenched
QCD. The analytical formulas for the average density of
the small Dirac eigenvalues are available \cite{SplitVerb2,akemann}.
They were first derived \cite{SplitVerb2} by exploiting the Toda
lattice hierarchy in the flavor index.
Comparisons of  random matrix predictions  \cite{SplitVerb2}
for the radial spectral density and lattice QCD results
\cite{tilopr,tilo-ger} 
are shown in the left panel of Fig. \ref{fig7}. In other cases, such as
the overlap Dirac operator \cite{bloch} and QCD with $N_c=2$ \cite{bittner},
a similar 
degree of agreement was found.
Both the spectral density and two-point correlations can be
derived from the Lagrangian  (\ref{lmu}), i.e. they 
are determined by two parameters, $F_\pi $ and $ \Sigma$. This
can be exploited to extract these low-energy constants. 
For example, $F_\pi$ and $ \Sigma$ were
determined  \cite{heller,OW} (see also \cite{heller2})
from the correlators shown in the two right panels
of Fig. \ref{fig7}.


\section{~Chiral~Symmetry~Breaking~at~{\colgy $\mu~\ne~0$}}

\begin{figure}[b!]
\begin{center}
\bmini{5.5cm} \vspace*{-0.8cm}
\includegraphics[width=4cm]{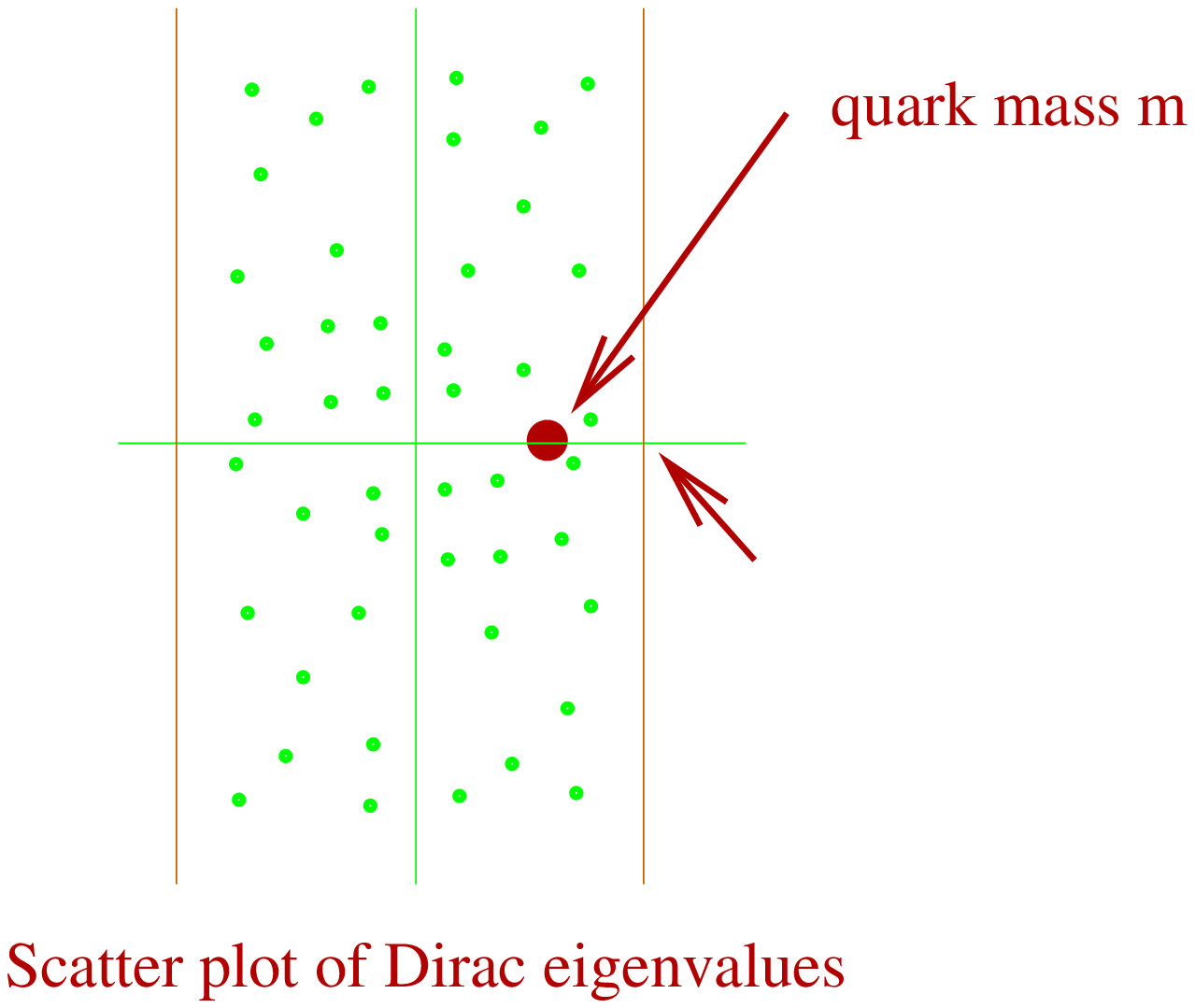}\hspace*{0.5cm}
\emini
\bmini{8cm}
\includegraphics[width=5.5cm]{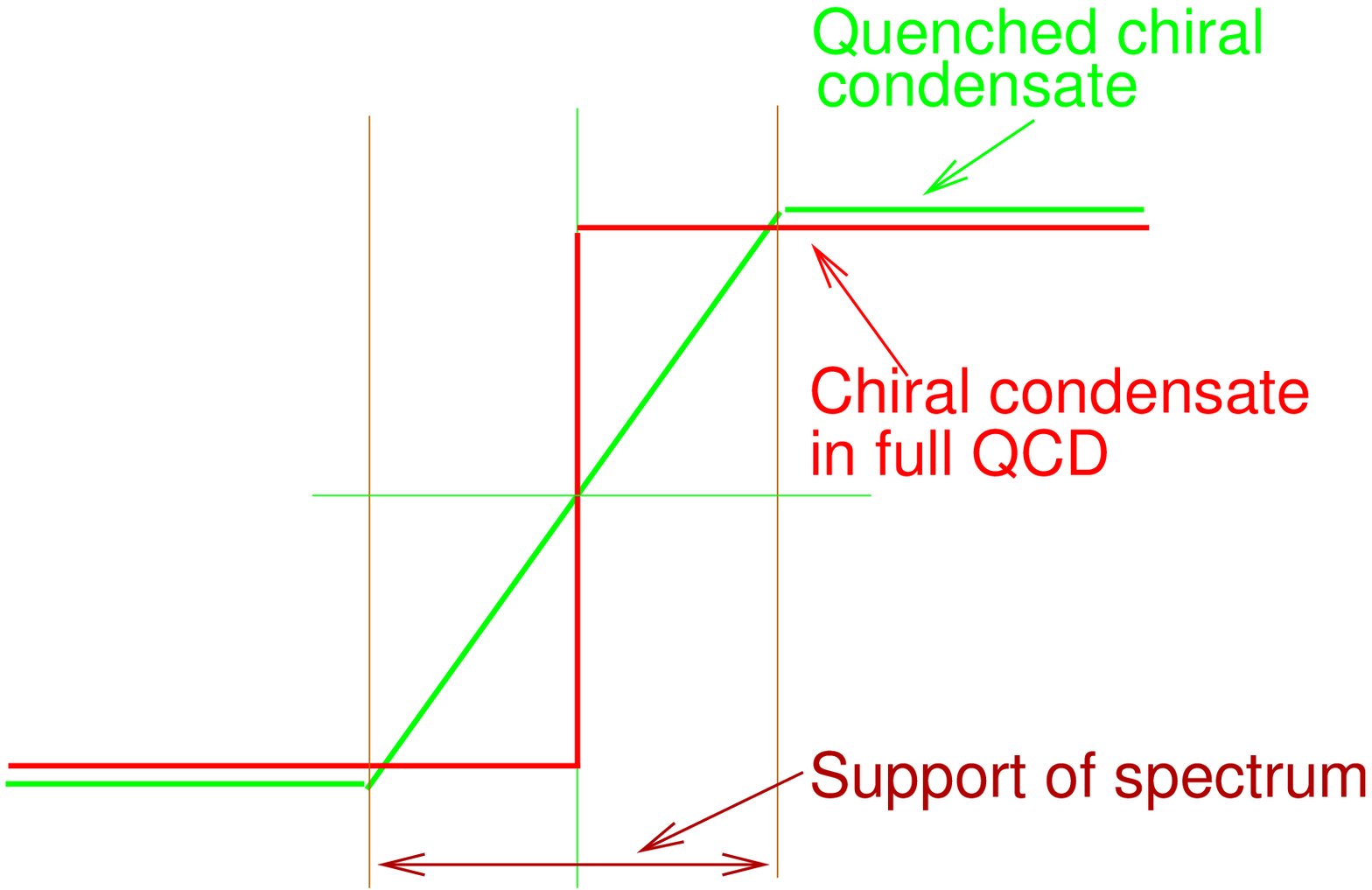}\\
\emini
\put(-11,-0.2){\colgy $\frac{\mu^2 F^2}{2\Sigma}$}
\put(-10.5,1){\colgy $\Sigma(m) =\frac 1V \sum_k \frac 1{m + \lambda_k}$}
\put(-4.3,-0.1){\colgy $m$}
\put(-3.3,1.1){\colgy $\Sigma(m)$}\\
\end{center}
\label{fig8}
\caption{Chiral condensate of quenched and full QCD.}
\end{figure}

The full QCD partition function at $\mu\neq0$ which is the average of
\be
\det(D+m+\mu\gamma_0) = |\det(D+m+\mu \gamma_0)| e^{i\theta}, 
\qquad \theta \ne 0,
\ee
has properties which are drastically different from the
phase quenched partition function where the phase factor is absent.
In particular, $\mu_c = m_N/3$ instead of $m_\pi/2$, so that
the free energy remains $\mu$-independent until $\mu=m_N/3$. 
For $\mu < m_N/3$ the chiral condensate remains 
discontinuous at $m=0$, whereas the chiral condensate of the
phase quenched theory approaches zero for $m\to 0$ (see Fig. \ref{fig8}).
The only difference between the phase quenched partition function and
the full QCD partition function is the phase of the fermion determinant.
We conclude that the phase factor is responsible for the discontinuity
of the chiral condensate.
How can this happen if for each configuration the support of the
spectrum is approximately the same? This problem  known as
the ``Silver Blaze Problem'' \cite{cohen} was solved in \cite{OSV}.

\subsection{Unquenched Spectral Density}

\begin{figure}[t!]
\begin{center}
\includegraphics[width=4.0cm]{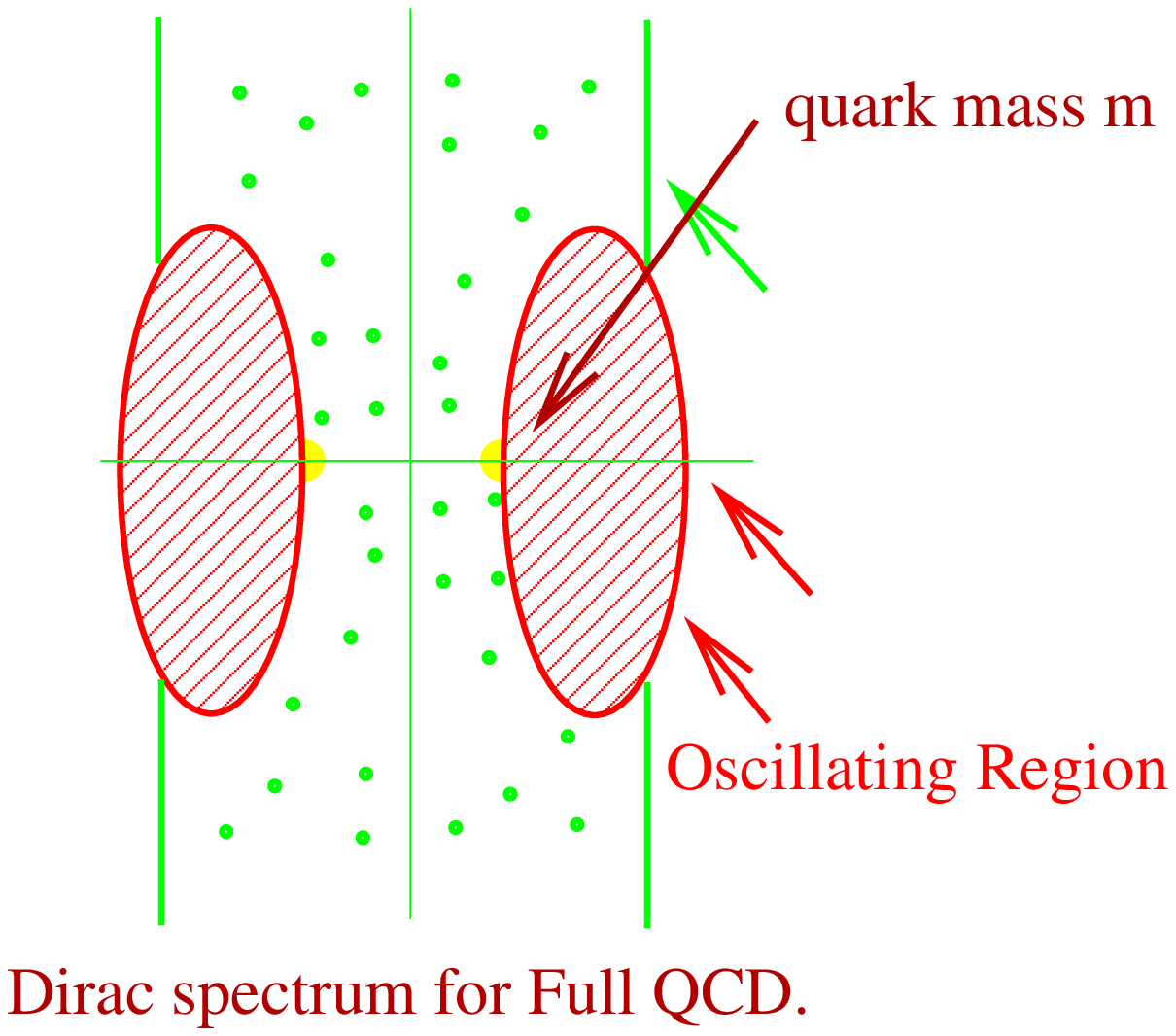}\hspace*{2cm}
\includegraphics[width=6.5cm]{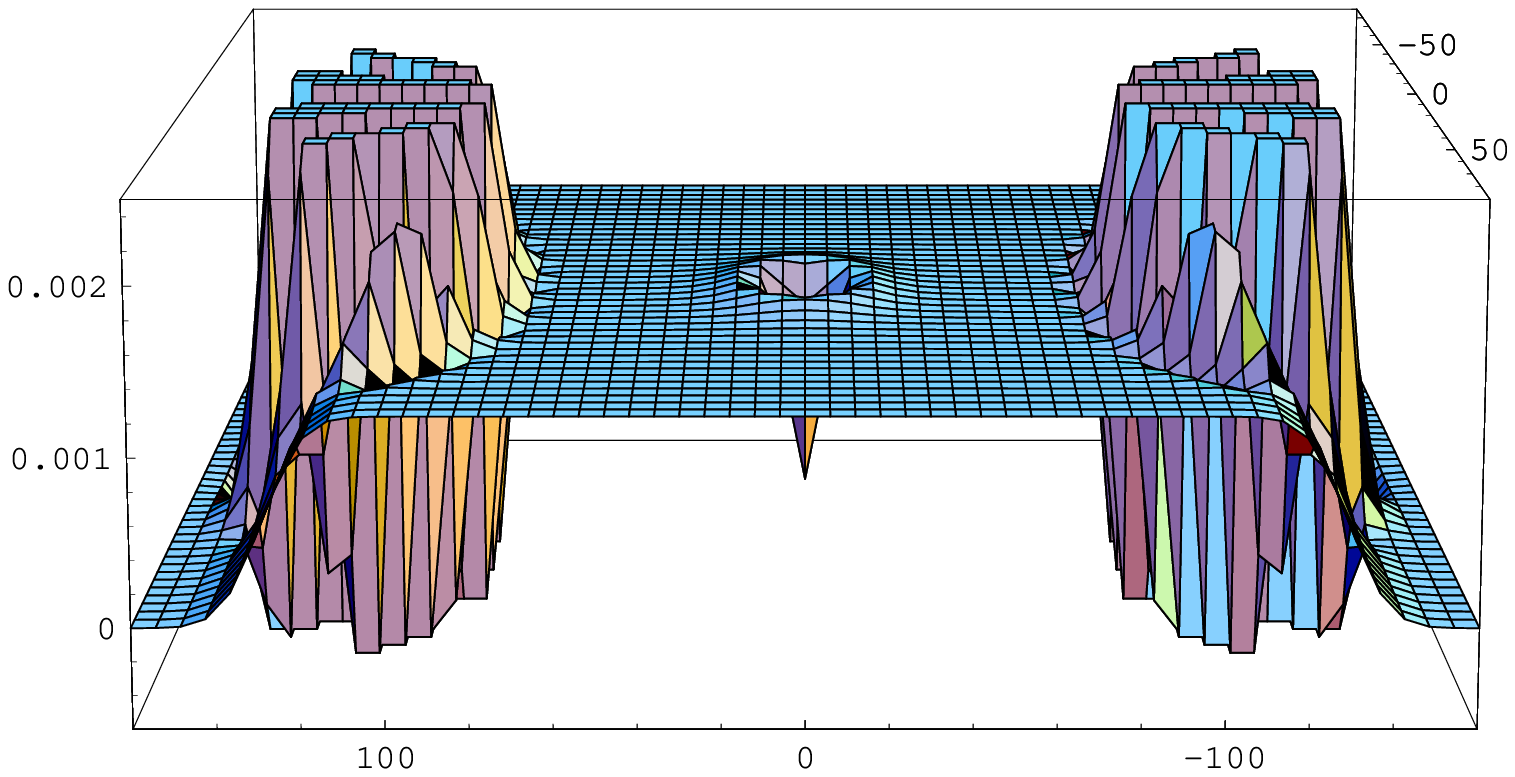} 
\put(-9.9,2.3){\green $\frac {2F^2\mu^2}{\Sigma}$}
\put(-9.8,1.3){\red $\frac 83 \frac {\mu^2F^2}{\Sigma} -\frac m3$}
\caption{Support (left) and real part (right, taken from \cite{splitrev}) 
of Dirac spectral density for QCD with $N_f=1$ and $\mu\ne 0$.}
\label{fig9}
\end{center}
\end{figure}

The spectral density for QCD with dynamical fermions is given by
\be
\rho_{N_f}(\lambda) = \langle \sum_k \delta^2 (\lambda-\lambda_k)
{\det}^{N_f}(D+m +\mu\gamma_0) 
 \rangle .
\ee
Because of the phase of the fermion determinant, this density is in general
complex and can be decomposed as
$
\rho_{N_f}(\lambda) = \rho_{N_f=0}(\lambda) + \rho_U(\lambda)\nn.
$
The chiral condensate can then be decomposed as
$
\Sigma_{N_f}(m) = \Sigma_{N_f=0}(m) + \Sigma_U(m),  
$
so that the discontinuity in {\colgy $\Sigma(m)$}  
is due to {\colgy $\rho_U$}.  
Asymptotically it behaves as \cite{AOSV}
\be{\colgy
\rho_U \sim e^{\frac 23 \mu^2 F^2 V} e^{\frac 23 i {\rm Im }(\lambda) \Sigma V}\nn}
\ee
and vanishes outside an ellips starting at {\colgy ${\rm Re }(\lambda) =m$}
(see Fig.~\ref{fig9}) \cite{OSV}. 
In the right part of this figure we show the
real part of the spectral density 
for QCD with one flavor at
nonzero chemical potential.

This result explains the mechanism of chiral symmetry breaking
at nonzero chemical potential.
 The phase of the
fermion determinant rotates the pion condensate back into a chiral 
condensate, but it does so in an unexpected way \cite{OSV}. The same 
mechanism is at play  for 1d QCD at $\mu \ne 0$ \cite{qcd1d}.

\section{ Phase of the Fermion Determinant}

The magnitude of the sign problem can be measured by means
of the expectation value of the phase factor of the fermion
determiant which can be defined in two  ways
\be
\langle e^{2i\theta} \rangle_{N_f} = \frac 1{Z_{N_f}}\left \langle 
\frac{\det(D+\mu\gamma_0+m)}{{\det}^*(D+\mu\gamma_0+m)}
{\det}^{N_f} (D+\mu \gamma_0+m)\right \rangle, \quad
\langle e^{2i\theta} \rangle_{1+1^* }
= \frac{Z_{N_f=2}}{Z_{1+1^*}}.\nn
\ee
The average $\langle\cdots \rangle$ is with respect to the Yang-Mills
action.
The sign problem is { managable}
 when the average phase factor remains
finite in the thermodynamic limit. 
In the microscopic domain it is possible to obtain exact
analytical expressions for the average phase factor by exploiting
the equivalence between QCD and RMT in this domain. 
\begin{figure}[b!]
\vspace*{-0.5cm}
\begin{center}
\includegraphics[width=4cm]{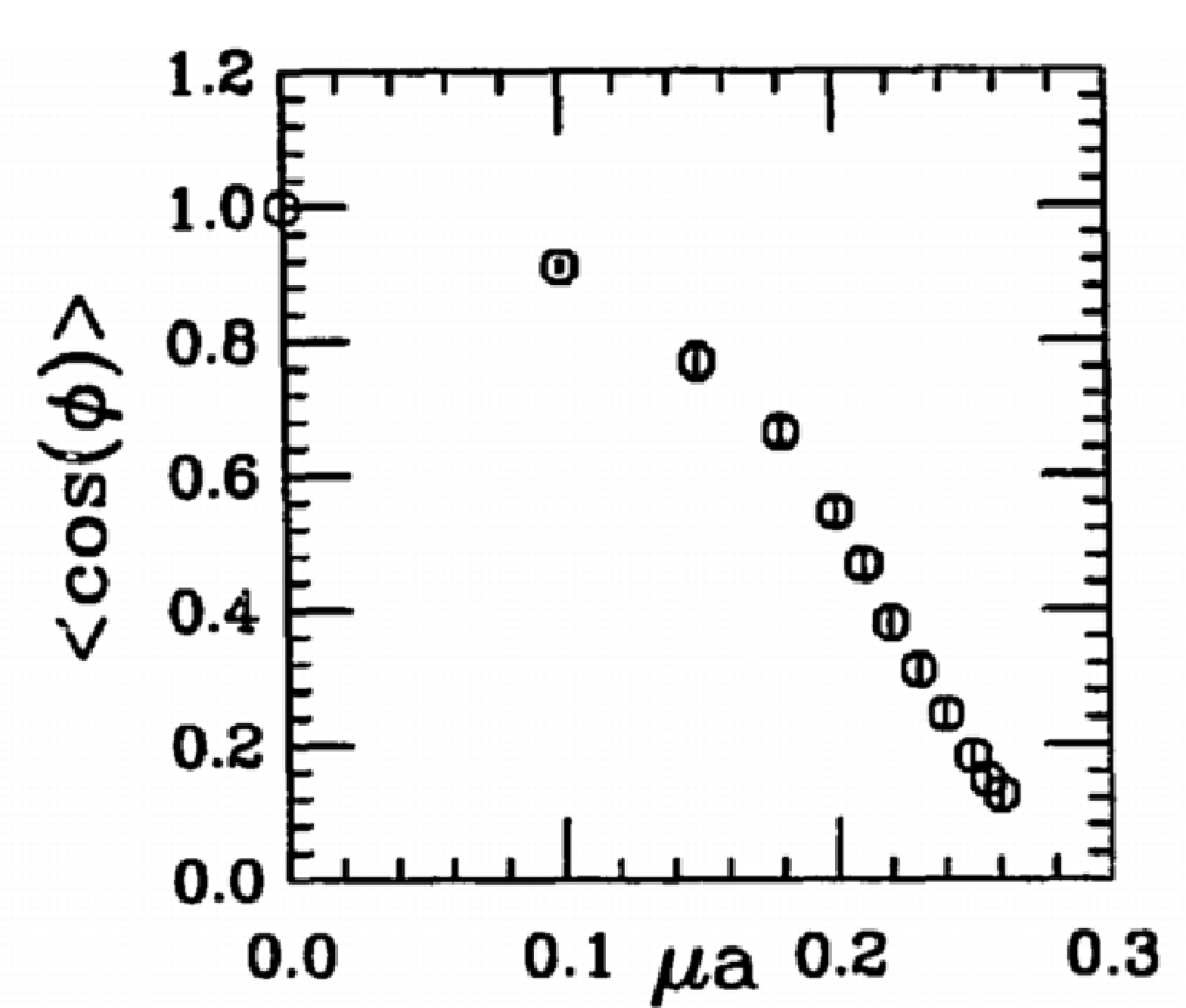}\hspace*{2cm}
{\includegraphics[width=4cm]{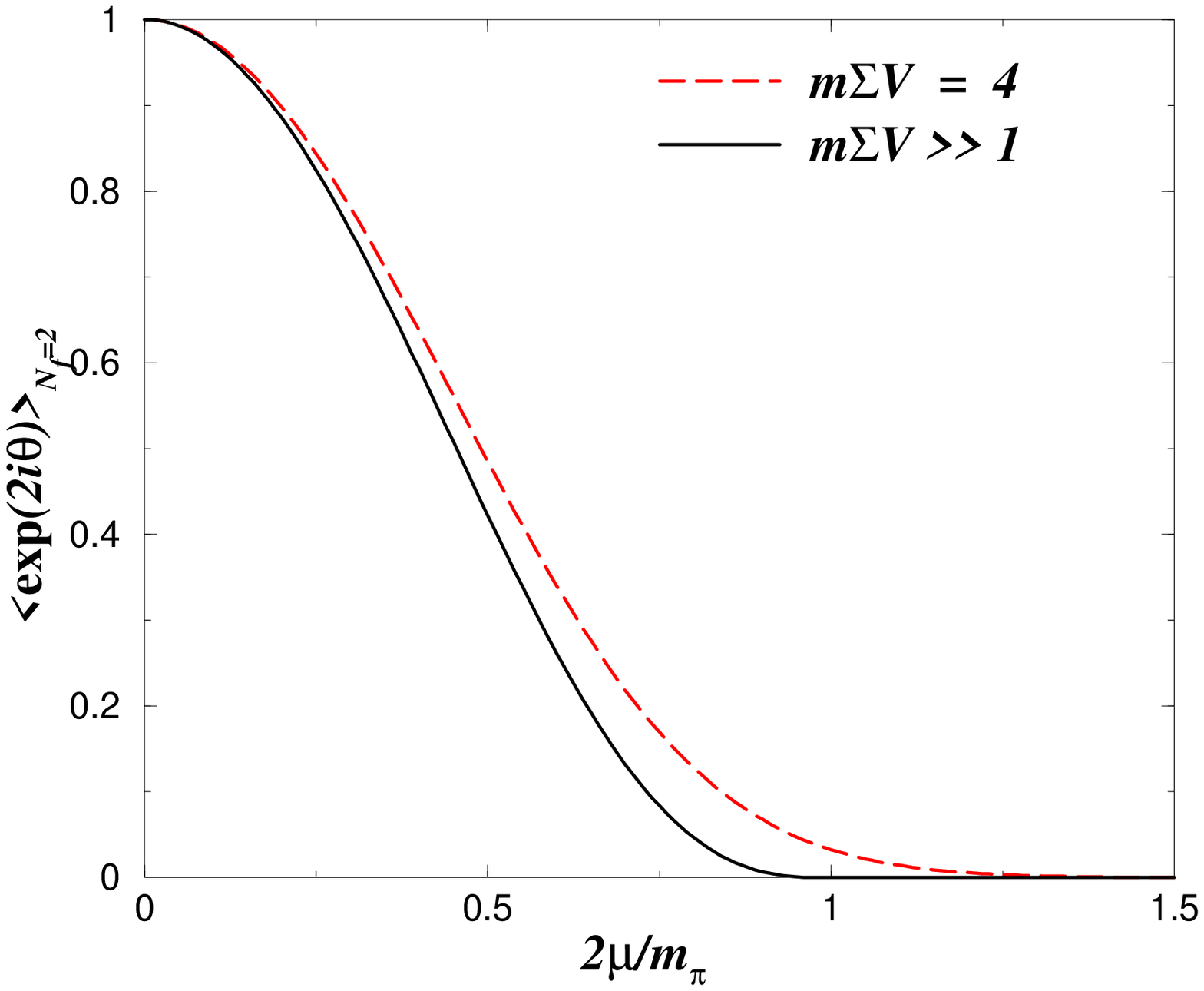}}
\end{center}
\caption{Average phase factor. Lattice QCD results are shown left
(taken from \cite{Toussaint}) and the exact microscopic result \cite{SV1}
is shown right. }
\label{fig10}
\end{figure}
For $\mu <m_\pi/2$ the free energy of both QCD and phase quenched QCD
are independent of $\mu$. This does not imply that the average phase
factor is $\mu$-independent. 
The  $\mu$-dependence originates from the charged Goldstone
bosons with mass $m_\pi \pm 2\mu$, and for $N_f$ flavors the mean field 
result \cite{SV1,SV2} for $\langle\exp(2i\theta)\rangle$  reads
$( 1 -  4\mu^2 /m_\pi^2 )^{N_f+1}$.
The exact result for the average phase factor for $N_f=2$ is shown in 
Fig. \ref{fig10} (right), where lattice results \cite{Toussaint} are
also shown (left). The exact result 
has an essential singularity at $\mu=0$, but its thermodyanmic limit
agrees with the mean result.

\section{ Conclusions}

The equivalence of chiral random matrix theory and QCD has been
exploited succesfully to derive a host of analytical results.
 Among others, eigenvalue fluctuations
predicted by chRMT have been observed in 
lattice simulations, the phases of QCD can be understood in terms of
spectral flow, observables can be extracted from the fluctuations of
the smallest eigenvalues, 
the sign problem is not serious when the quark mass is outside
the domain of the eigenvalues, and mean field results can be obtained
from random matrix theory. Summarizing, 
chiral random matrix theory is a powerful 
tool for analyzing the infrared domain of QCD.

\section*{Acknowledgements}

The YITP is thanked for its hospitality.
G. Akemann, J. Osborn and P.H. Damgaard are acknowledged 
for valuable discussions. This work was
supported  by US DOE Grant No. DE-FG-88ER40388 (JV), 
the Villum Kann Rasmussen Foundation (JV), the Danish National Bank (JV) and 
the Carslberg Foundation (KS).


\end{document}